# Enhanced Hydrogen Bonding to Water Can Explain the Outstanding Solubility of β-D-Glucose in Water

Imre Bakó[a], László Pusztai[b,c], Szilvia Pothoczki[b*]

[a] HUN-REN Research Centre for Natural Sciences, H-1117 Budapest, Magyar tudósok körútja 2., Hungary
[b] HUN-REN Wigner Research Centre for Physics, H-1121 Budapest, Konkoly-Thege M. út 29-33., Hungary
[c] International Research Organization for Advanced Science and Technology (IROAST), Kumamoto University, 2-39-1 Kurokami, Chuo-ku, Kumamoto, 860-8555, Japan

*Correspondence to: pothoczki.szilvia@wigner.hun-ren.hu

**Abstract**

*Ab initio* molecular dynamics (AIMD) simulations have been performed on aqueous solutions of four simple sugars, α-D-glucose, β-D-glucose, α-D-mannose and α-D-galactose. Hydrogen bonding (HB) properties, such as the number of donor and acceptor type HB-s, and the lengths and strengths of hydrogen bonds between sugar and water molecules, have been determined. Related electronic properties, such as the dipole moments of water molecules and partial charges of the sugar O-atoms, have also been calculated. The hydrophilic and hydrophobic shells were characterized by means of spatial distribution functions. β-D-glucose has been found to form the highest number of hydrophilic and the smallest number of hydrophobic connections to neighboring water molecules. The average sugar-water H-bond length was the shortest for β-D-glucose, which suggests that these are the strongest such H-bonds. Furthermore, β-D-glucose appears to stand out in terms of symmetry properties of both its hydrophilic and hydrophobic hydration shells. In summary, in all aspects considered here, there seems to be a correlation between the distinct characteristics of β-D-glucose and its outstanding solubility in water.

**Introduction**

Carbohydrates are one of the most crucial biomolecules, which play a principal role in several biological processes such as molecular recognition, structural stabilization, and modification of proteins and nucleic acids. They also act as cryoprotective molecules in living cells [1]. Additionally, they play a significant role in many industrial applications related to, e.g., the food, biotechnology, biofuels, and cosmetics industries. [1-3] The most important, and therefore the most studied, of them are hexopyranose sugars that have six different ring conformations. Their five chirality centers lead to 32 diastereoisomers, characterized by the axial or equatorial orientation of ring substituents.

Many important properties derive from the conformational flexibility and hydroxymethyl and exocyclic hydroxyl group orientations, and their interaction with water. The interconversion between conformers is hindered by high free-energy barriers, leading to characteristic times of the order of hundred picoseconds to microseconds [4,5]. This long time is one of the problems while constructing appropriate force fields for molecular dynamics (MD) or Monte Carlo (MC) simulations. [4-6]

Although carbohydrates are generally considered hydrophilic compounds, they have substantial hydrophobicity that varies with their structure. In the gas phase, the energetically most stable conformations are the ones whose hydroxyl groups form a well-defined intramolecular H-bonded pattern. [7,8] The competition between these intramolecular H-bonds and the intermolecular ones that are formed between the oxygens of sugars and water molecules, together with hydrophobic interactions, determine the solvation shell of these molecules.

Various theoretical [9-22] and experimental techniques [23-33] have been used to explore and understand the relationship between the conformational and configurational properties of carbohydrates, and the structural arrangement of water molecules in the first hydration shell at the atomistic level. Approximately 10 to 12 water molecules are bonded to a central hexopyranose monosaccharide molecule via H-bonds according to classical and *ab initio* MD simulations, and neutron diffraction results. [20,21,23,24] Additionally, 8 to 10 water molecules are attached through significantly weaker interactions to the hydrophobic parts ('surface') of the sugar molecules. [23,24] The energetic properties of these water molecules are determined through the van der Waals and CH…$O_{water}$ type interactions. Therefore, it is important to consider the polarizability and charge transfer properties of the alcoholic groups of sugar and water molecules when describing interactions between them.

Here, we consider four simple sugars, α-D-glucose, β-D-glucose, α-D-galactose, and α-D-mannose that possess very similar molecular structures (see Figure 1) but quite different water solubilities (see Table 1). A major focus of the present work is to find whether this property may be related to the hydrogen bonding environment of sugar molecules. It is, however, important to note here that the H-bond interaction is only one of the important factors associated with solubility. Other types of interactions, e.g. changes in the magnitude of the water-water interactions and changes in the hydration sphere of water molecules, induced by the presence of solute molecules, might well also play important roles in the process.

Along the lines drawn just above, i.e., by examining hydrogen bonding down to the electronic structure level, we hope to be able to see and demonstrate differences between the four sugars in question that lead to an understanding of, among others, their different solubilities. Further, our extensive studies of the hydration shell of these monosaccharides, based on ab initio molecular dynamics simulations (AIMD, see, e.g., Ref. [20]), may provide information concerning why these molecules can recognize some specific patterns on a macromolecule. The main advantage of AIMD over classical computer simulations is that AIMD is able to take charge transfer and polarization effects that arise from sugar-water interactions, into account.

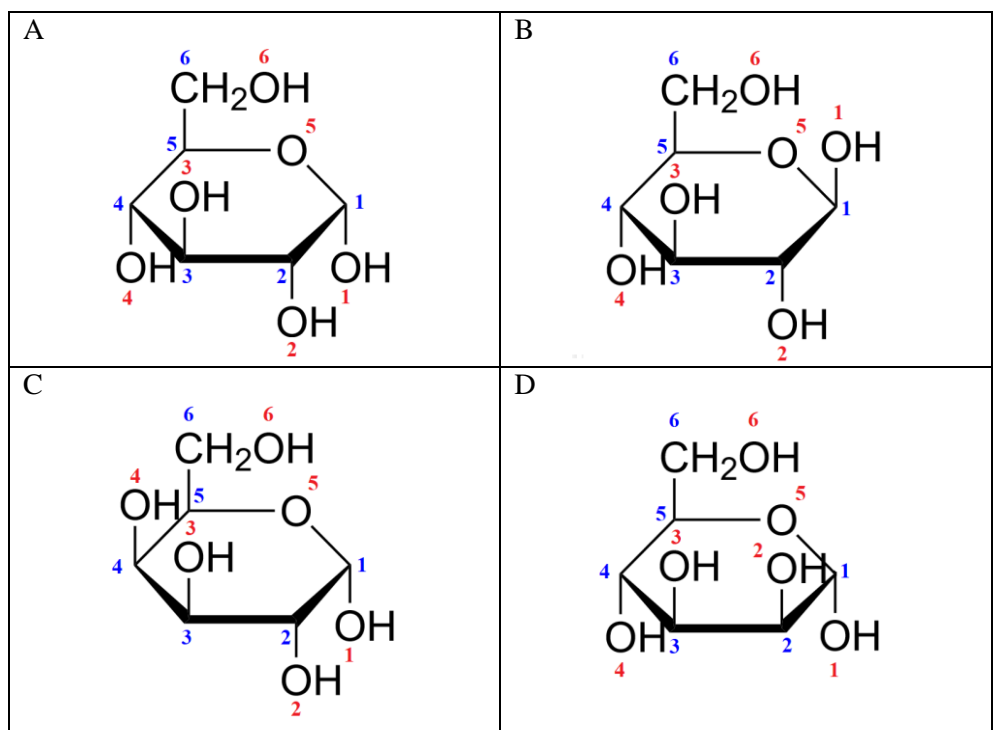

Figure 1. Haworth projections of (A) α-D-glucose, (B) β-D-glucose, (C) α-D-galactose, (D) α-D-mannose.

Table 1. Water solubilities of the hexopyranose monosaccharides considered here. (https://pubchem.ncbi.nlm.nih.gov/)

| α-D-glucose | β-D-glucose | α-D-galactose | α-D-mannose |
|---|---|---|---|
| 500 mg/ml | 1200 mg/ml | 683 mg/ml | 713 mg/ml |

**Computational details**

The initial particle configurations were generated from classical MD simulations using the GROMACS software. [34] For monosaccharide molecules the Charmm36 all-atom interaction potential [12], for the water molecules the SPC/E [35] explicit water model was used. Periodic boundary conditions were applied and the box size was 1.52577 nm (α-glucose), 1.52640 nm (β-glucose), 1.51058 nm (α-galactose) and 1.51170 nm (α-mannose) containing one monosaccharide molecule and 99 water molecules for a total of 321 atoms in each case. The average temperature of the four systems during the NVT simulations was 320 K using the Berendsen thermostat [36]. The Newtonian equations of motions were integrated via the leapfrog algorithm, using a time step of 1 fs. The total run time was 300 ns. The particle-mesh Ewald [37,38] algorithm was used for handling the long-range electrostatic forces and potentials.

The present results arise from DFT-based (BLYP/D3) ab initio molecular dynamics simulations that have been performed at 320 K in a periodic setup using the CP2K [39] ab initio molecular dynamics code. The norm-conserving Goedeker-Teter-Hutter pseudopotentials were used. We employed a triple zeta basis set with double polarization (TZV2P). The plane-wave basis sets using 300 Ry a charge density cut-off in the CP2K program. The time-step was 0.5 fs. Particle configurations were collected for subsequent analyses from 100 ps runs after 40 ps of equilibration.

The Wannier centers (necessary for the molecule dipole moment calculations, see, e.g., Ref. [34]) were collected in every 50th time-step (25 fs) in the trajectories of the production run. Fractional atomic charges were derived from the Bader-type analysis, using Henkelman's code [40] and analyzed every 100[th] time-step (every 50 fs).

Following our earlier work, [41] two molecules were considered to be hydrogen bonded to each other when they were at a distance r(O···H) < 2.5 Å and the H-O…O angle was < 30°. The limiting distance for in the definition is based on the well-defined first intermolecular minima of the corresponding partial radial distribution functions between the sugar and water hydroxyl groups. (c.f. Fig. S3) On the other hand, there are water molecules, which do not bond through H-bonds to the central sugar molecule, but they form a van der Waals complex with the sugar molecule. Those water molecules, for which it is satisfied that $r(C_i \cdots O_{water}) < 4.5$ Å (c.f. Fig. S4) are considered as member of the hydrophobic shell.

**Results and discussion**

*Average coordination numbers*

The total average coordination numbers can be taken as the sum of numbers of hydrophilic (N$_{HPHILE}$) and hydrophobic (N$_{HPHOBE}$) components. The hydrophilic part incorporates all water molecules that are bonded to a central sugar molecule through an H-bond. Results from the present AIMD simulations, provided in Table 2, show that β-D-glucose molecules form the most H-bonds with the surrounding water molecules, while in the other three studied cases, the number of H-bonds differs only a little.

Concerning the hydrophobic shell, water molecules for which $r(C_i \ldots O_{water}) < 4.5$ Å are taken into account. It is β-D-glucose that coordinates the smallest number of water molecules by hydrophobic interactions (see Table 2), whereas roughly the same number of non-hydrogen-bonded water molecules can be found in the first hydration shell of the other three sugar molecules. As far as we are aware, this property has not been investigated for these systems previously. It should be noted that the accuracy of the values in Table 2 (calculated using the bootstrapping method [42,43]) for both the hydrophilic and hydrophobic hydration spheres was found to be within a few hundredths.

Table 2. Average coordination numbers for the four hexopyranose monosaccharides investigated here from the CP2K simulation

|   | α-D-glucose | β-D-glucose | α-D-galactose | α-D-mannose |
| --- | --- | --- | --- | --- |
| N$_{HPHILE}$ | 10.83 | 11.41 | 10.73 | 10.88 |
| N$_{HPHOBE}$ | 9.22 | 8.47 | 9.39 | 9.17 |

*H-bonding properties*

The hydrogen bonding ability of the different oxygen sites (for the assignation, c.f. Fig. 1) is shown in Figure 2a. There are five hydroxyl groups in the monosaccharide molecules studied here, and all of them can act as a donor of one H-bond and as an acceptor of two H-bonds. Additionally, the ring O atom can form two H-bonds with water, as acceptor. In all cases, the donor coordination number is larger than 0.9 but it stays below 1: these values are only slightly different among the sugar molecules considered here. The smallest average coordination number (approx. 0.81) belongs to the O2 donor site of α-D-mannose, that is somewhat out of line. Concerning acceptor properties, the O6 sites were found to be the most attractive and the O2 site is the second most occupied. Again, α-D-mannose exhibits a slightly different behavior, as the best acceptor was found the O3 site, and the second best is O6. For each studied case, the ring oxygens have proven to be the poorest acceptors.

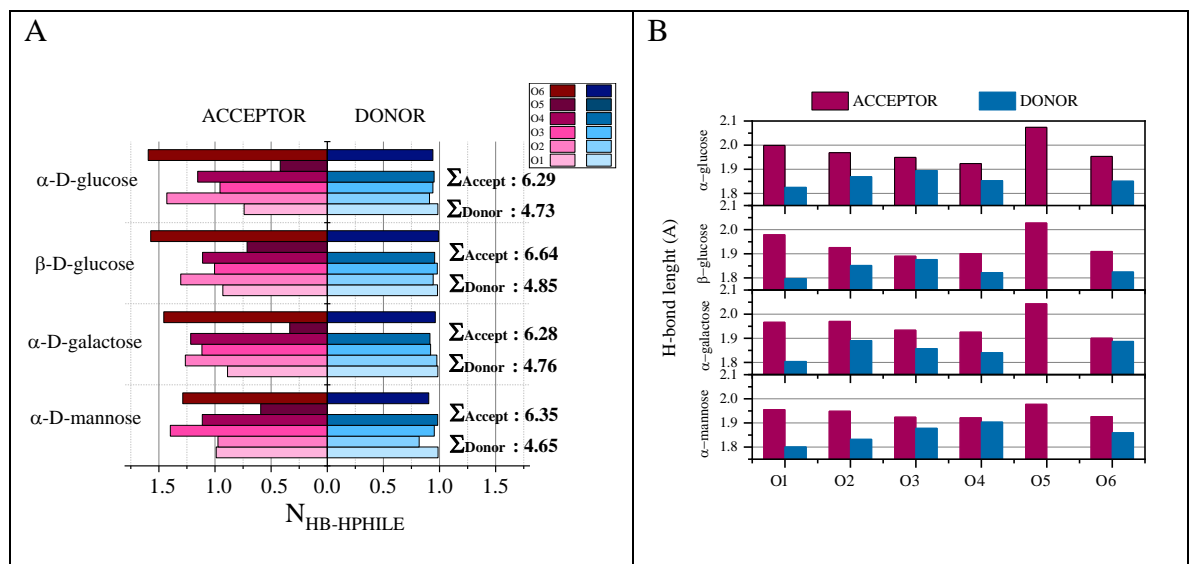

Figure 2. (A) The average number of acceptor and donor H-bonds. (B) The average length of acceptor and donor H-bonds.

Overall, β-D-glucose forms the highest number of H-bonds both as an acceptor and as a donor (see Figure 2). Another significant difference between the H-bonding abilities of different sugar molecules is the extent of involvement in H-bonding as acceptor (see Figure 2a). For each of the sugar molecules we have studied, the O5 oxygen atom has the fewest water molecules attached to it by H-bonds. One finds the largest number of water molecules that H-bonds to the O5 position in β-D-glucose: this, in terms of H-bonding properties, is arguably the most significant difference between the monosaccharides studied here. Note also that the number of hydrogen bonds formed by sugars as donors and acceptors is slightly greater than the number of water molecules present in the first hydrophilic sphere, suggesting that some water molecules form H-bonds with sugars as both donors and acceptors (see Table 2 vs. Figure 2a).

The strengths of the donor and acceptor types of H-bonds can be estimated by scrutinizing $H_i...O_{water}$ and $O_i...H_{water}$ distances (Fig. 2b), where $H_i$ and $O_i$, the hydrogen and oxygen atoms of the hydroxyl group connected to the i-th C atoms (c.f. Fig. 1). It is found that the donor distances are significantly smaller than the acceptor ones: this implies that H-bond donor interactions are stronger than acceptor ones. The longest average H-bond acceptor distance was detected for the $O5..H_wO$ type H-bond in α-D-glucose. The largest differences between donor and acceptor H-bonded distances have always been detected at the O1 sites (see Fig. 2b). The shortest average donor H-bond length of a sugar molecule was revealed in β-D-glucose, while the longest one, related to a weakest H-bond strength, was found in α-D-glucose. In agreement with this, the average H-bond length was also the shortest in the β-D-glucose solution.

We note that Suzuki et al. [21] using a different approach, namely the Wannier projection method [42], made a similar observation. They established that the $H_i...O_{water}$ (sugar-water) bond is shorter, and the $O_i…H_{water}$ (sugar-water) bond is longer, than the same type of H-bond between two water molecules in the bulk. Beyond the calculation of related characteristic distances, we also show in Supporting Information that water-water interactions are highly dependent on the number of hydrogen-bonded neighboring molecules (c.f. Figure S5), that is, on the coordination numbers (c.f. Table 2, Fig 2a).

*Electronic properties*

Concerning electronic properties, the Wannier dipole moment of water molecules that are H-bonded directly to a monosaccharide are displayed in Figure 3: results are classified according to the H-bonded coordination number of water molecules. Water molecules, regardless of which sugar oxygen atom they are connected to, mostly are fourfold-coordinated (one neighbor is the sugar and the other three are water molecules). It is found that the dipole moment of fourfold-coordinated water molecules is larger than that of threefold-coordinated ones: this is the same behaviour as found for molecules in pure water [44]. That is, the dipole moment of water molecules, again (cf. also Ref. [44]), clearly depends on their H-bonded neighborhood.

Furthermore, the acceptor or donor nature of monosaccharide molecules was also taken into account during this classification. Comparing the two glucose systems, no clear difference can be detected in terms of the behavior of "contact" water molecules that are located in the first shell of glucose molecules.

An interesting observation, however, is in order here: water molecules acting as (H-) donors (to the saccharide molecule) possess, on average, larger dipole moments than the ones with acceptor roles. Although this point, not being vital from the point of view of the main focus of the present work, is not pursued any further, the conjecture is worth making that this behaviour occurs in parallel (or perhaps, even as a consequence) with that donor H-bonds are shorter (and thus, stronger) than acceptor ones (cf. Figure 2).

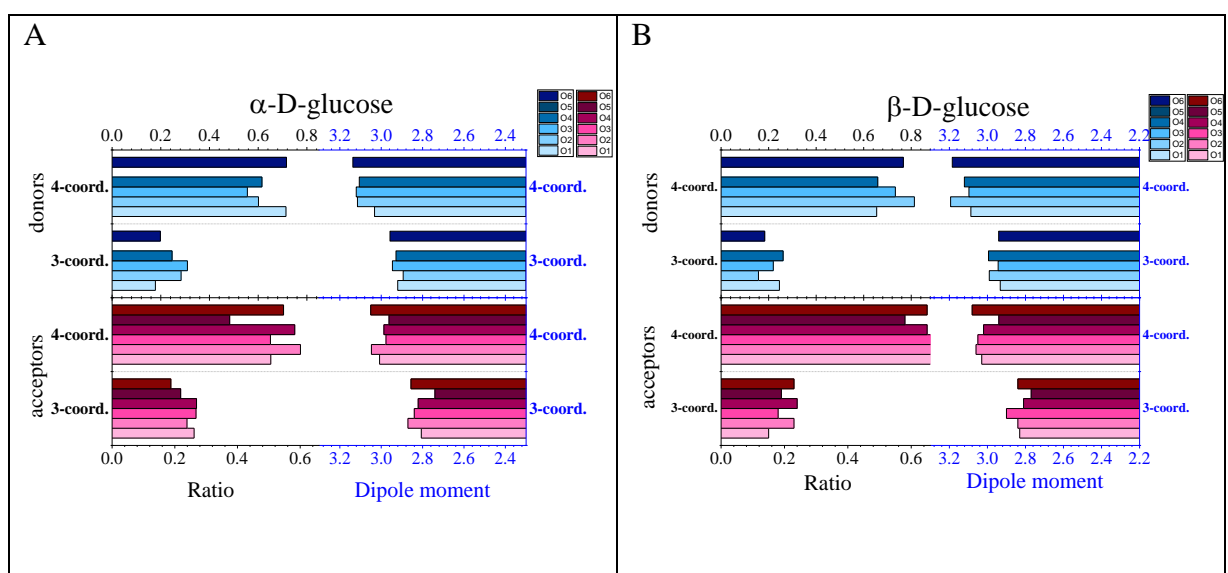

Figure 3. (A) Properties of three and four-coordinated water molecules around α-D-glucose (B) Properties of three and four-coordinated water molecules around β-D-glucose.

Turning our attention to the electronic properties of the monosaccharide molecules, Bader charges [45] of OH groups and O5 atoms in the ring have been calculated. As seen in Table 3, the charges belonging to O2H, O3H, O4H, O6H are almost identical for each investigated sugar. On the other hand, charges on the anomeric O1H groups are significantly different: they are the most negative, and among the sugars here, β-D-glucose sticks out with the largest negative value.

Table 3. Bader charges /-e/ on the OH sites of sugar molecules in the solutions.

|              | O1H    | O2H    | O3H    | O4H    | O6H    |
|--------------|--------|--------|--------|--------|--------|
| α-D-glucose  | -0.669 | -0.539 | -0.544 | -0.564 | -0.551 |
| β-D-glucose  | -0.727 | -0.568 | -0.573 | -0.570 | -0.564 |
| α-D-galactose| -0.625 | -0.546 | -0.567 | -0.547 | -0.547 |
| α-D-mannose  | -0.650 | -0.554 | -0.558 | -0.565 | -0.551 |

A similar conclusion can be drawn if charges on the O atoms in the sugar molecules are considered. The charge distribution on the O atoms is also most pronounced on the O1 atom as shown by the dipole moment data in Table 4. The smallest negative charge is on the O5 atom, which is clearly related to the fact that the O5…$H_w$ H-bond is always the longest. Additionally, the total charge of each monosaccharide studied here in aqueous solutions is slightly negative: -0.022 $e$ (α-D-glucose), -0.018 $e$ (β-D-glucose), -0.017 $e$ (α-D-galactose and α-D-mannose). This small negative charge is the result of a rather complex electron transfer process, since sugar forms H-bonds with more water molecules as an acceptor than as a donor. Accordingly, sugar molecules should possess a positive charge. However, the length of the H-bonds is significantly shorter when the sugar molecule acts as a donor. During the formation of stronger H-bonds, there is a greater electron transfer from water molecules to sugar molecules than vice versa. (Note that this effect would not be visible in classical simulations.)

Table 4. Bader charges /-e/ on the O atoms in the ring of sugar molecules in the solutions.

|              | O1     | O2     | O3     | O4     | O5     | O6     |
|--------------|--------|--------|--------|--------|--------|--------|
| α-D-glucose  | -1.316 | -1.206 | -1.197 | -1.232 | -1.054 | -1.209 |
| β-D-glucose  | -1.407 | -1.230 | -1.243 | -1.237 | -1.064 | -1.220 |
| α-D-galactose| -1.273 | -1.200 | -1.222 | -1.203 | -1.067 | -1.200 |
| α-D-mannose  | -1.324 | -1.207 | -1.216 | -1.213 | -1.104 | -1.200 |

The magnitude of the charge (Table 4) and the dipole moment (Table 5) connected to the O atoms characterize the asymmetry of the charge distribution within the monosaccharide molecules: both the charge and the dipole moment are always largest for the O1 oxygen. Furthermore, both values are the highest for β-D-glucose among the monosaccharides studied here.

Table 5. Dipole moment /D/ of the O atoms in the ring of sugar molecules in the solutions.

|  | O1 | O2 | O3 | O4 | O5 | O6 |
|---|---|---|---|---|---|---|
| α-D-glucose | 0.507 | 0.391 | 0.373 | 0.411 | 0.240 | 0.386 |
| β-D-glucose | 0.625 | 0.389 | 0.433 | 0.405 | 0.247 | 0.378 |
| α-D-galactose | 0.462 | 0.389 | 0.416 | 0.367 | 0.265 | 0.382 |
| α-D-mannose | 0.525 | 0.395 | 0.390 | 0.368 | 0.312 | 0.377 |

*Hydration sphere of monosaccharides*

Spatial distribution functions [46] (SDF) around central sugar molecules have also been calculated. SDFs are able to provide unique 3D representations of hydration shells (hydrophilic and hydrophobic) around a central monosaccharide molecule that are the consequences of the different orientations of hydroxyl groups. The origin of the coordinate system was the center of mass (CM) of the sugar, the XY plane is defined by the positions of $C_5$, CM and $C_1$, and the +x direction is defined as the bisector of the $C_5$—CM—$C_1$ angle. Figure 4a-c displays the SDF for α-D-glucose, and Figure 4d-f for β-D-glucose, as typical examples.

The red and yellow colors denote the hydrophilic and the hydrophobic shells, respectively. Note that the average density of $O_{water}$ atoms around a monosaccharide molecule is higher than the bulk density of water (0.033 molecules/Å$^3$). Water molecules are located in well-identifiable regions around the sugar: in the directions of hydroxyl groups (along the $O_{sugar}$-$H_{sugar}$ vector), and the lone pairs of $O_{sugar}$. In hydrophobic shells, these regions are mainly located in the directions of $C_{sugar}$-$H_{sugar}$.

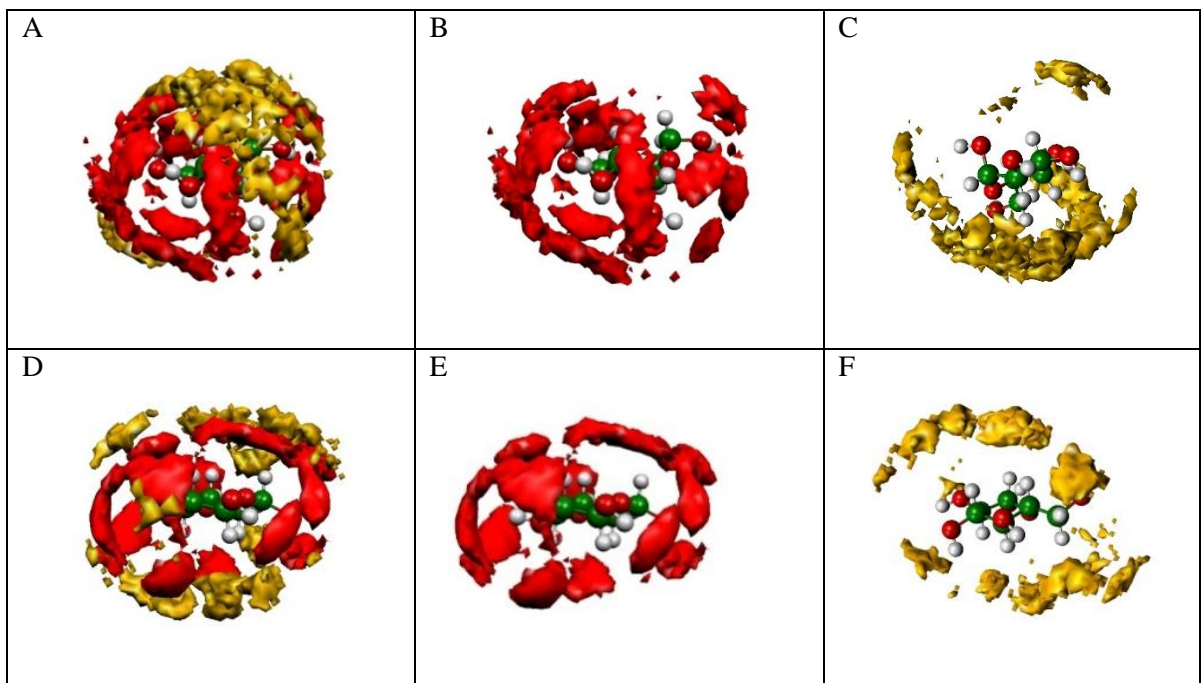

Figure 4. Spatial distribution functions around α-D-glucose (upper panels) and β-D-glucose (lower panels). (A) and (D) Hydrophobic and hydrophilic shells together, (B) and (E) hydrophilic shell only, (C) and (F) hydrophobic shell only.

In order to extract quantitative information from the SDF, we determined the number of water molecules below and above the XY plane of the sugar molecules for both types of hydration spheres (Figure 5). In general, there is a significantly larger number of water molecules in the hydrophobic shell above the plane: β-D-glucose is one of the most symmetric in this respect. On the other hand, the hydrophilic hydration sphere of each, but of β-D-glucose, monosaccharide contains significantly more molecules below the plane. This is related to the direction of the OH group (O1) attached to the first carbon atom (C1): for α anomers the −$CH_2OH$ (C6 and O6) group lies on the opposite side of the XY plane of the molecule, while for the β anomer, O1 and O6 are on the same side (c.f. Fig.1). It may be conjectured then that these asymmetric features of the hydration shell can play a significant role in molecular recognition. It should be noted that the accuracy of the values in Figure 5 (calculated, again, using the bootstrapping method [42, 43]) was found to be within a few hundredths.

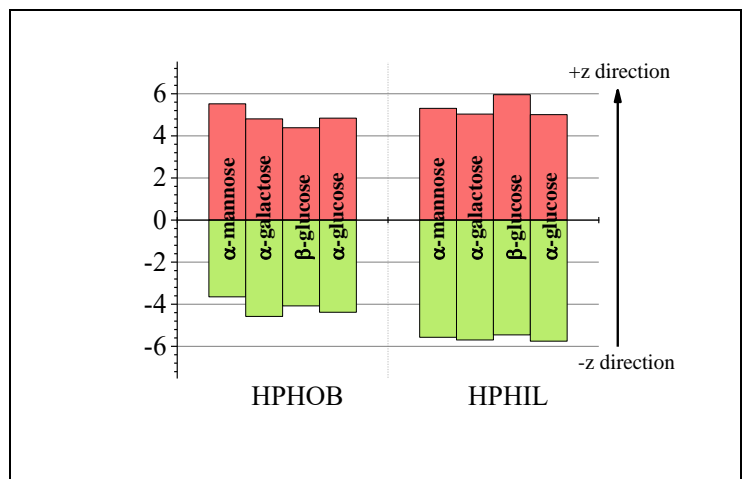

Figure 5 The number of hydrophobic and hydrophilic H-bonded water molecules above and below the XY molecular plane of monosaccharide molecules considered here.

**Conclusions**

In summary, *ab initio* molecular dynamics simulations have been performed for aqueous solutions of four simple sugars, α-D-glucose, β-D-glucose, α-D-mannose and α-D-galactose.

(1) It has been established that β-D-glucose forms the highest, while α-D-glucose forms one of the smallest, number of hydrophilic hydrogen bonds with surrounding water molecules. Concerning hydrophobic interactions with nearby water molecules, β-D-glucose has the smallest number of water bounded by hydrophobically. These findings are in accord with the fact that β-D-glucose is the most, whereas α-D-glucose is the least soluble in water.

(2) The largest number of water molecules that H-bonds to the O5 position is found in the aqueous solution of β-D-glucose.

(3) The average H-bond length when the OH group of the sugar molecule is involved as a donor in H-bonds was the shortest/longest for β-D-glucose/α-D-glucose, respectively. In agreement with this, the average H-bond length was also the shortest for β-D-glucose.

(4) The Bader charges derived from AIMD simulations for oxygen atoms at positions 2,3,4 and 6 are very similar for all monosaccharides studied. On the other hand, the O atom at position 1 has a significantly more negative charge. Of all the O atoms, the one at position 5 has the least negative charge.

(5) Scrutinizing the spatial structure of the hydration spheres of sugar molecules has revealed apparent asymmetries of the hydrophobic shell, whereas the hydrophilic shell is much more symmetric. β-D-glucose appears to be outstanding in terms of both of its hydrophilic and hydrophobic hydration shells.

Acknowledgment

The authors greatly appreciate helpful discussions with Dr. László Jicsinszky (University of Turin). The authors are grateful to the National Research, Development and Innovation Office (NRDIO (NKFIH), Hungary) for financial support via grants Nos. 142429 and FK 128656.

**References:**

[1] R. A. Dwek, Glycobiology: Toward Understanding the Function of Sugars, Chem. Rev. 96 (1996) 683−720, https://doi.org/10.1021/cr940283b.

[2] C. Clarke, R. J. Woods, J. Gluska, A. Cooper, M. A. Nutley, G. J. Boons, Involvement of water in carbohydrate-protein binding, J. Am. Chem. Soc. 123 (2001) 12238–12247, https://doi.org/10.1021/ja004315q.

[3] H. J. Zhu, D. Liu, Vy P. Tran, Z. Wu, K. Jiang, H. Zhu, J. Zhang, C. Gibbons, B. Xue, H. Shi, P. G. Wang, N-Linked Glycosylation Prevents Deamidation of Glycopeptide and Glycoprotein, ACS Chem. Biol. 15 (2020) 3197-3205, https://doi.org/10.1021/acschembio.0c00734.

[4] B. L. Foley, M. B. Tessier, R. J. Woods, Carbohydrate Force Fields, WIREs Comput. Mol. Sci. 2 (2012) 652−697, https://doi.org/10.1002/wcms.89.

[5] H. S. Hansen, P. H. Hünenberger, A reoptimized GROMOS force field for hexopyranose-based carbohydrates accounting for the relative free energies of ring conformers, anomers, epimers, hydroxymethyl rotamers, and glycosidic linkage conformers, J Comput. Chem. 32 (2011) 998–1032, https://doi.org/10.1002/jcc.21675.

[6] W. Plazinski, A. Lonardi, P. H. Hünenberger, Revision of the GROMOS 56A6CARBO Force Field: Improving the Description of Ring-Conformational Equilibria in Hexopyranose-Based Carbohydrates Chains, J. Comput. Chem. 37 (2016) 354–365, https://doi.org/10.1002/jcc.24229.

[7] I. Tvaroška, J. P. Carver, Ab Initio Molecular Orbital Calculation of Carbohydrate Model Compounds 6. The Gauche Effect and Conformations of the Hydroxymethyl and Methoxymethyl Groups, J. Phys. Chem. B 101 (1997) 2992−2999, https://doi.org/10.1021/jp963766n.

[8] I. Tvaroška, F. R. Taravel, J. P. Utille, J. P. Carver, Quantum Mechanical and NMR Spectroscopy Studies on the Conformations of the Hydroxymethyl and Methoxymethyl Groups in Aldohexosides, Carbohydr. Res. 337 (2002) 353−367, https://doi.org/10.1016/S0008-6215(01)00315-9.

[9] V. Krautler, M. Muller, P. H. Hunenberger, Conformation, Dynamics, Solvation and Relative Stabilities of Selected β-Hexopyranoses in Water: A Molecular Dynamics Study with the GROMOs 45A4 Force Field, Carbohydr. Res. 342, (2007) 2097−2124, https://doi.org/10.1016/j.carres.2007.05.011.

[10] C. Chen, W. Z. Li, Y. C. Song, L. D. Weng, N. Zhang, Formation of Water and Glucose Clusters by Hydrogen Bonds in Glucose Aqueous Solutions, Comput. Theor. Chem. 984 (2012) 85−92, https://doi.org/10.1016/j.comptc.2012.01.013.

[11] K. N. Kirschner, A. B. Yongye, S. M. Tschampel, J. González-Outeiriño, C. R. Daniels, B. L. Foley, R. J. Woods, GLYCAM06: A Generalizable Biomolecular Force Field. Carbohydrates, J. Comput. Chem. 29 (2008) 622−655, https://doi.org/10.1002/jcc.20820.

[12] O. Guvench, S. N. Greene, G. Kamath, J. W. Brady, R. M. Venable, R. W. Pastor, A. D. Mackerell, Additive empirical force field for hexopyranose monosaccharides, J Comput. Chem. 29 (2008) 2543–2564, https://doi.org/10.1002/jcc.21004.

[13] L. Pol-Fachin, V. H. Rusu, H. Verli, R. D. Lins, GROMOS 53A6GLYC, an improved GROMOS force field for hexopyranose based carbohydrates, J. Chem. Theory Comput. 8 (2012) 4681–4690, https://doi.org/10.1021/ct300479h.

[14] H. Senderowitz, C. Parish, W. C. Still, Carbohydrates: united Atom AMBER* parameterization of pyranoses and simulations yielding anomeric free energies, J. Am. Chem. Soc. 118 (1996) 2078–2086, https://doi.org/10.1021/ja9529652.

[15] P. Pandey, S. S. Mallajosyula, Influence of Polarization on Carbohydrate Hydration: A Comparative Study Using Additive and Polarizable Force Fields J. Phys. Chem. B 120 (2016) 6621−6633, https://doi.org/10.1021/acs.jpcb.6b05546.

[16] D. S. Patel, X. He, A. D. Jr. MacKerell, Polarizable Empirical Force Field for Hexopyranose Monosaccharides Based on the Classical Drude Oscillator, J. Phys. Chem. B 119 (2015) 637−652, https://doi.org/10.1021/jp412696m.

[17] J. L. Dashnau, K. A. Sharp, J. M. Vanderkooi, Carbohydrate Intramolecular Hydrogen Bonding Cooperativity and Its Effect on Water Structure, J. Phys. Chem. B 109 (2005) 24152-24159, https://doi.org/10.1021/jp0543072.

[18] U. Schnupf, J. L. Willett, F. Momany, DFTMD Studies of Glucose and Epimers: Anomeric Ratios, Rotamer Populations, and Hydration Energies, Carbohydr. Res. 345 (2010) 503−511, https://doi.org/10.1016/j.carres.2009.12.001.

[19] M. Hoffmann, J. Rychlewski, Effects of Substituting a OH Group by a F Atom in D-Glucose. Ab Initio and DFT Analysis, J. Am. Chem. Soc. 123 (2001) 2308−2316, https://doi.org/10.1021/ja003198w.

[20] C. Molteni, M. Parrinello, Glucose in aqueous solution by first principles molecular dynamics ,J. Am. Chem. Soc. 120 (1998) 2168−2171, https://doi.org/10.1021/ja973008q.

[21] T. Suzuki, The Local Configurations in Sugar−Water Hydrogen Bonds, Phys. Chem. Chem. Phys. 10 (2008) 96−105, https://doi.org/10.1039/B708719E.

[22] K. Tomobe, E. Yamamoto, D. Kojić, Y. Sato, M. Yasui, K. Yasuoka, Origin of the blueshift of water molecules at interfaces of hydrophilic cyclic compounds, Sci. Adv. (2017) e1701400, DOI: 10.1126/sciadv.1701400.

[23] P. E. Mason, G. W. Neilson, J. E. Enderby, M.-L. Saboungi, J. W. Brady, Structure of Aqueous Glucose Solutions as Determined by Neutron Diffraction with Isotopic Substitution Experiments and Molecular Dynamics Calculations, J. Phys. Chem. B 109, (2005) 13104-13111, https://doi.org/10.1021/jp040622x.

[24] P. E. Mason, G. W. Neilson, J. E. Enderby, M.-L. Saboungi, G. Cuello, J. W. Brady, Neutron diffraction and simulation studies of the exocyclic hydroxymethyl conformation of glucose, J. Chem. Phys. 125 (2006) 224505, https://doi.org/10.1021/ja051376l.

[25] H. Rhys, F. Bruni, S. Imberti, S. E. McLain, M. Ricci, Glucose and Mannose: A Link between Hydration and Sweetness, J. Phys. Chem. B 121 (2017) 7771−7776, https://doi.org/10.1021/acs.jpcb.7b03919.

[26] F. Bruni, C. Di Mino, S. Imberti, S. E. McLain, N. H. Rhys, M. Ricci, Hydrogen Bond Length as a Key To Understanding Sweetness, J. Phys. Chem. Lett. 9 (2018) 3667−3672, https://doi.org/10.1021/acs.jpclett.8b01280.

[27] S. Imberti, S. E. McLain, N. H. Rhys, F. Bruni, M. Ricci, Role of Water in Sucrose, Lactose, and Sucralose Taste: The Sweeter, The Wetter?, ACS Omega 4 (2019) 22392−22398, https://doi.org/10.1021/acsomega.9b02794.

[28] K. Shiraga, T. Suzuki, N. Kondo, T. Tajima, M. Nakamura, H. Togo, A. Hirata, K. Ajito, Y. Ogawa, Broadband dielectric spectroscopy of glucose aqueous solution: Analysis of the

hydration state and the hydrogen bond network, J. Chem. Phys. 142 (2015) 234504, https://doi.org/10.1063/1.4922482.

[29] M. Heyden, E. Bründermann, U. Heugen, G. Niehues, D. Leitner, M. Havenith, Long-Range Influence of Carbohydrates on the Solvation Dynamics of Water Answers from Terahertz Absorption Measurements and Molecular Modeling Simulations, J. Am. Chem. Soc. 130 (2008) 5773−5779, https://doi.org/10.1021/ja0781083.

[30] U. Heugen, G. Schwaab, E. Bründermann, M. Heyden, X. Yu, D. Leitner, M. Havenith, Solute-Induced Retardation of Water Dynamics Probed Directly by Terahertz Spectroscopy, Proc. Natl. Acad. Sci. U.S.A. 103 (2006) 12301−12306, https://doi.org/10.1073/pnas.0604897103.

[31] S. A. Galema, H. Hoiland, Stereochemical aspects of hydration of carbohydrates in aqueous solutions. 3. Density and ultrasound measurements, J. Phys. Chem. 95 (1991) 5321-5326, https://doi.org/10.1021/j100166a073.

[32] Y. Nishida, H. Hori, H. Ohrui, H. Meguro, 1H NMR analyses of rotameric distribution of C5–C6 bonds of d-glucopyranoses in solution, J. Carbohydr. Chem. 7 (1988) 239–250, https://doi.org/10.1080/07328308808058917.

[33] J. Ø. Duus, C. H. Gotfredsen, K. Bock, Carbohydrate Structural Determination by NMR Spectroscopy: Modern Methods and Limitations, Chem. Rev. 100 (2000) 4589−4614, https://doi.org/10.1021/cr990302n.

[34] D. van der Spoel, E. Lindahl, B. Hess, G. Groenhof, A. E. Mark. H. J. Berendsen, GROMACS: fast. flexible. and free, J. Comp. Chem. 26 (2005) 1701, https://doi.org/10.1002/jcc.20291.

[35] H. J. C. Berendsen, J. R. Grigera, T. P. Straatsma, The missing term in effective pair potentials, J. Phys. Chem. 91 (1987) 6269-6271, https://doi.org/10.1021/j100308a038.

[36] H. J. C. Berendsen, J. P. M. Postma, A. DiNola, J. R. Haak, Molecular dynamics with coupling to an external bath, J. Chem. Phys. 81 (1984) 3684, https://doi.org/10.1063/1.448118.

[37] U. Essman, L. Perera, M. L. Berkowitz, T. Darden, H. Lee, L. G. Pedersen, A smooth particle mesh Ewald method, J. Chem. Phys. 103 (1995) 8577, https://doi.org/10.1063/1.470117.

[38] T. Darden, D. York, L. Pedersen, Particle mesh Ewald: An N·log(N) method for Ewald sums in large systems, J. Chem. Phys. 98 (1993) 10089, https://doi.org/10.1063/1.464397.

[39] J. Hutter, M. Iannuzzi, F. Schiffmann, V. Vondele, cp2k: atomistic simulations of condensed matter systems, WIREs Comput. Mol. Sci. 4 (2014) 15, https://doi.org/10.1002/wcms.1159.

[40] G. Henkelman, A. Arnaldsson, H. Jónsson, A fast and robust algorithm for Bader decomposition of charge density. Comput. Mater. Sci. 36 (2006) 354-360, https://doi.org/10.1016/j.commatsci.2005.04.010.

[41] Sz. Pothoczki, I. Pethes, L. Pusztai, L. Temleitner, K. Ohara, I. Bakó, Properties of hydrogen bonded network in ethanol-water liquid mixtures as a function of temperature: diffraction experiments and computer simulations, J. Phys. Chem. B 125 (2021) 6272–6279, https://doi.org/10.1021/acs.jpcb.1c03122.

[42] B. Efron, Bootstrap Methods: Another Look at the Jackknife, Ann. Stat. 7 (1979) 1, http://www.jstor.org/stable/2958830.

[43] R. Beran, The Impact of the Bootstrap on Statistical Algorithms and Theory, Stat. Sci. 18 (2003) 175, DOI: 10.1214/ss/1063994972.

[44] I. Bakó, J. Daru, Sz. Pothoczki, L. Pusztai, K. Hermansson, Effects of H-bond asymmetry on the electronic properties of liquid water – An AIMD analysis, J. Mol. Liq. 293 (2019) 111579, https://doi.org/10.1016/j.molliq.2019.111579.

[45] R. F. W. Bader, C. F. Matta, Atomic Charges Are Measurable Quantum Expectation Values: A Rebuttal of Criticisms of QTAIM Charges, J. Phys. Chem. A 108 (2004) 8385, https://doi.org/10.1021/jp0482666.

[46] I. M. Svishchev, P. G. Kusalik, Structure in liquid water: A study of spatial distribution functions, J. Chem. Phys. 99 (1993) 3049-3058, https://doi.org/10.1063/1.465158.

# Supporting Information for

# Enhanced Hydrogen Bonding to Water Can Explain the Outstanding Solubility of β-D-Glucose in Water

Imre Bakó[a], László Pusztai[b,c,] Szilvia Pothoczki[b*]

[a] HUN-REN Research Centre for Natural Sciences, H-1117 Budapest, Magyar tudósok körútja 2., Hungary

[b] HUN-REN Wigner Research Centre for Physics, H-1121 Budapest, Konkoly-Thege M. út 29-33., Hungary

[c] International Research Organization for Advanced Science and Technology (IROAST), Kumamoto University, 2-39-1 Kurokami, Chuo-ku, Kumamoto, 860-8555, Japan

*Correspondence to: pothoczki.szilvia@wigner.hun-ren.hu

**Table of Contents:**

1. Comparison of classical and ab initio MD results

2. Dynamical properties

3. Conformational analyses

4. Partial radial distribution functions

5. Cooperative behavior of water molecules around the sugar molecule

1. Comparison of classical and ab initio MD results

A well-known shortcoming of AIMD is that it cannot explore the entire conformational space in practice, for simple computational considerations: the time needed for sugar molecules to change conformation is prohibitively long. The question is that whether this restriction has a significant effect on the present findings. In order to clarify the issue, all calculations that characterize H-bond properties have been repeated from analogous classical MD simulations (i.e. the same number of molecules and the same box length), exactly the way they are presented in the main text. Results for the classical MD simulations, including the average length (and corresponding standard deviations) of acceptor and donor H-bonds and the average coordination numbers in the hydrophilic and the hydrophobic shell, are given in Tables S1-S2.

**Table S1**. Average lengths (and corresponding standard deviations) of acceptor and donor H-bonds from classical Molecular Dynamics simulations (in Å).

|    | α-D-glucose | | | | β-D-glucose | | | |
|----|------|------|------|------|------|------|------|------|
|    | donor | $\sigma^2$ | acceptor | $\sigma^2$ | donor | $\sigma^2$ | acceptor | $\sigma^2$ |
| O1 | 2.023 | 0.194 | 2.010 | 0.195 | 2.017 | 0.193 | 2.017 | 0.191 |
| O2 | 2.030 | 0.194 | 1.993 | 0.228 | 2.027 | 0.195 | 1.994 | 0.228 |
| O3 | 2.025 | 0.192 | 1.979 | 0.192 | 2.024 | 0.193 | 1.991 | 0.195 |
| O4 | 2.017 | 0.192 | 1.977 | 0.192 | 2.016 | 0.193 | 1.978 | 0.190 |
| O5 |       |       | 1.988 | 0.193 |       |       | 1.981 | 0.193 |
| O6 | 2.046 | 0.196 | 1.974 | 0.182 | 2.044 | 0.197 | 1.971 | 0.179 |
|    | α-D-mannose | | | | α-D-galactose | | | |
|    | donor | $\sigma^2$ | acceptor | $\sigma^2$ | donor | $\sigma^2$ | acceptor | $\sigma^2$ |
| O1 | 2.017 | 0.193 | 2.017 | 0.191 | 2.028 | 0.195 | 2.007 | 0.195 |
| O2 | 2.027 | 0.195 | 1.991 | 0.195 | 2.031 | 0.195 | 1.989 | 0.192 |
| O3 | 2.024 | 0.193 | 1.978 | 0.190 | 2.041 | 0.196 | 1.976 | 0.186 |
| O4 | 2.016 | 0.193 | 1.981 | 0.193 | 2.029 | 0.194 | 1.983 | 0.191 |
| O5 |       |       | 1.994 | 0.228 |       |       | 1.968 | 0.229 |
| O6 | 2.044 | 0.197 | 1.971 | 0.179 | 2.028 | 0.195 | 1.969 | 0.181 |

**Table S2**. Average coordination numbers in the hydrophilic and the hydrophobic shell from classical MD simulations.

|  |    | α-D-glucose | β-D-glucose | α-D-galactose | α-D-mannose |
|---|----|------|------|------|------|
| $N_{HPHILE}$ | +z | 3.78 | 4.38 | 3.98 | 4.15 |
|              | -z | 4.23 | 3.87 | 4.03 | 4.04 |
|              | Σ  | **8.01** | **8.25** | **8.01** | **8.19** |
| $N_{HPHOBE}$ | +z | 5.38 | 4.73 | 5.63 | 5.27 |
|              | -z | 4.09 | 4.56 | 4.80 | 4.64 |
|              | Σ  | **9.47** | **9.29** | **10.43** | **9.91** |

In classical MD simulations, donor distances were found to be longer than acceptor ones, by a few hundredths of an Å on average. In AIMD, the opposite occurs (cf. Fig. 2b and Table S1), and differences are much larger. Arguably, it is the AIMD results that must be taken as more reliable, since classical MD cannot account for the charge transfer and polarization effects adequately, and these have a significant influence on the property in question. Our main finding, the tendency regarding the symmetric–asymmetric shape of the hydration spheres can also be detected by classical MD simulations (c.f. Figure 5 and Table S2). That is, this phenomenon is not related to the problem that AIMD is unable to explore the full conformational space of monosaccharide molecules in solutions.

2. Dynamical properties

Dynamical properties, such as self-diffusion constants for both the monosaccharide and water molecules, reorientation times of OH vectors of water molecules in solutions, and lifetimes of H-bonds, have been calculated from AIMD simulation (see Table S3-S5). These data show clearly that our simulations have, indeed, been conducted in the liquid phase.

**Table S3**. Self-diffusion constant calculated by the Einstein relation, without taking PBC effects into account.

| Systems | Self-diffusion constant ($10^{-8}$ m$^2$/s) |
|---|---|
| α-D-glucose | 0.125 |
| β-D-glucose | 0.117 |
| α-D-galactose | 0.101 |
| α-D-mannose | 0.127 |
| water | 0.160 |

**Table S4.** The characteristic time of reorientation of OH vectors of water molecules in the studied monosaccharide solutions (in ps).

|  | P1(OH) (ps) | P2(OH) (ps) | P HOH plane (ps) |
|---|---|---|---|
| α-D-glucose | 16.13 | 8.77 | 12.50 |
| β-D-glucose | 22.22 | 13.16 | 17.86 |
| α-D-galactose | 18.87 | 12.50 | 14.29 |
| α-D-mannose | 20.83 | 10.87 | 13.16 |

**Table S5**. Lifetimes of H-bonds calculated from autocorrelation functions (in ps).

| Lifetime (ps) | HB (water-water) | HB (sugar-water) |
|---|---|---|
| α-D-glucose | 2.85 | 5.09 |
| β-D-glucose | 3.68 | 9.06 |
| α-D-galactose | 3.34 | 4.29 |
| α-D-mannose | 3.04 | 5.43 |

3. Conformational analyses

Pyranose rings remained in their $^4C_1$ state during the simulations. A recent publication [1] revealed that the $^4C_1$ conformation is stable even in a very long simulation (5 μs). The authors showed that, according to Protein Data Bank, more than 80 % of the molecules are in $^4C_1$ configuration for the monosaccharides investigated here. [2] To present detailed conformational profile O5-C5-C6-O6, O5-C5-C6-O6 and C2-C1-O1-H1 dihedrals are shown in Figs. S1 and S2. In the classical MD simulations, the force field defined dihedral angle distributions were detected, with reasonable agreement with the experimentally determined abundance. In the AIMD simulations, α- and β-D glucose and α-D-galactose molecules, with respect to the hydroxymethyl group, are in the gauche-trans state, whereas about 17% of the saccharide molecules were found in the gauche-gauche conformation in the α-D-mannose solution.

Figure S1 and S2 show the conformation analyses in the classical MD (left-hand panels) and the ab initio MD simulations (right-hand panels) for the four studied monosaccharides. In the classical MD simulations, we reproduced the literature conformer ratio. [2] Note that in ab initio MD simulations the time scale does not allow for the sugar molecules to turn from one conformational state to another.

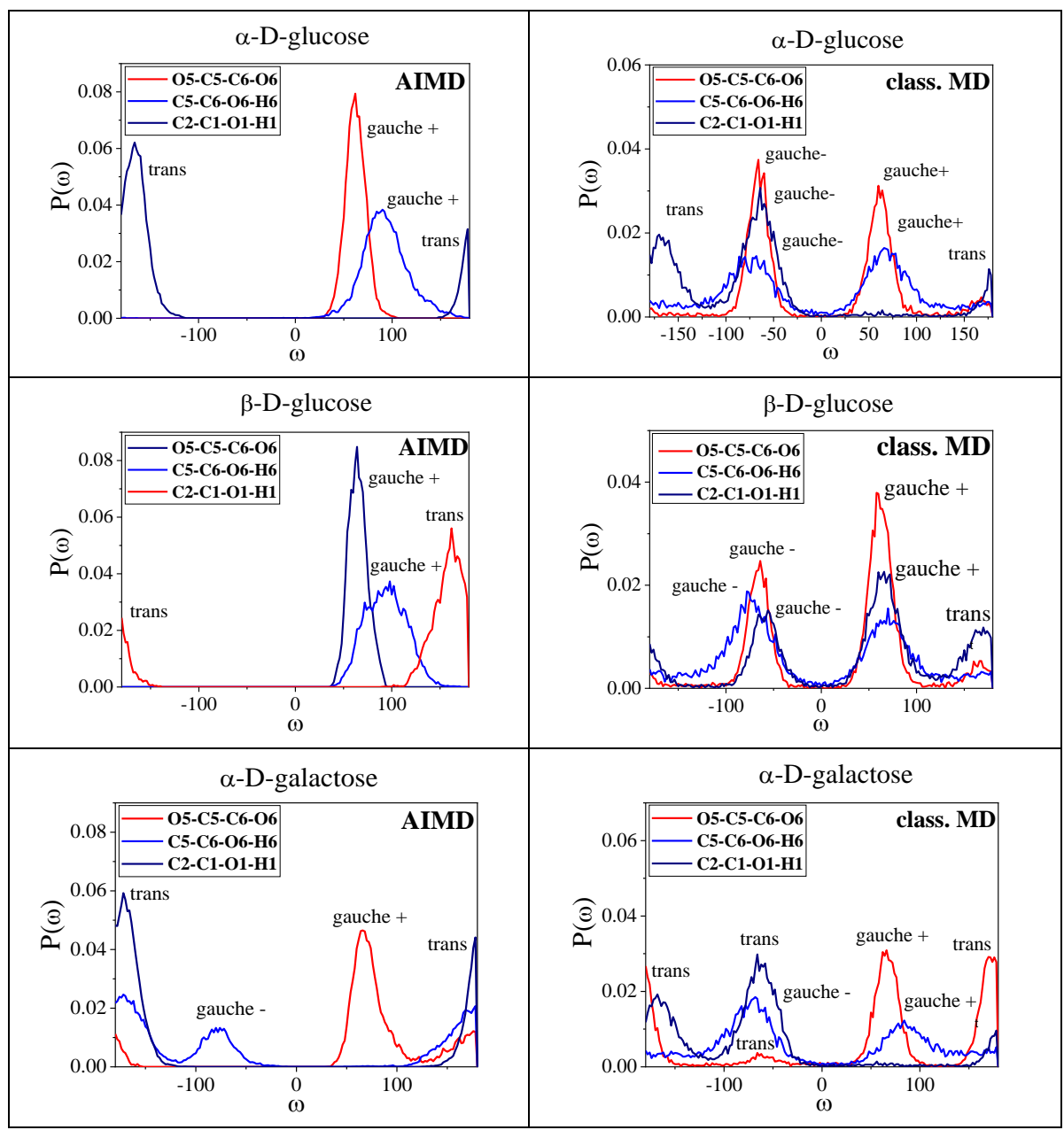

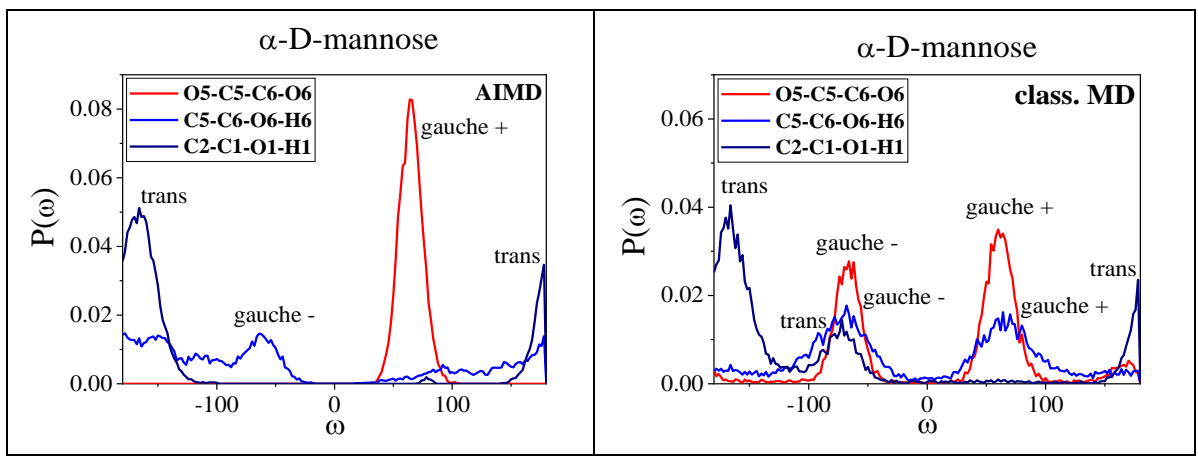

**Figure S1.** Characteristic dihedral angles.

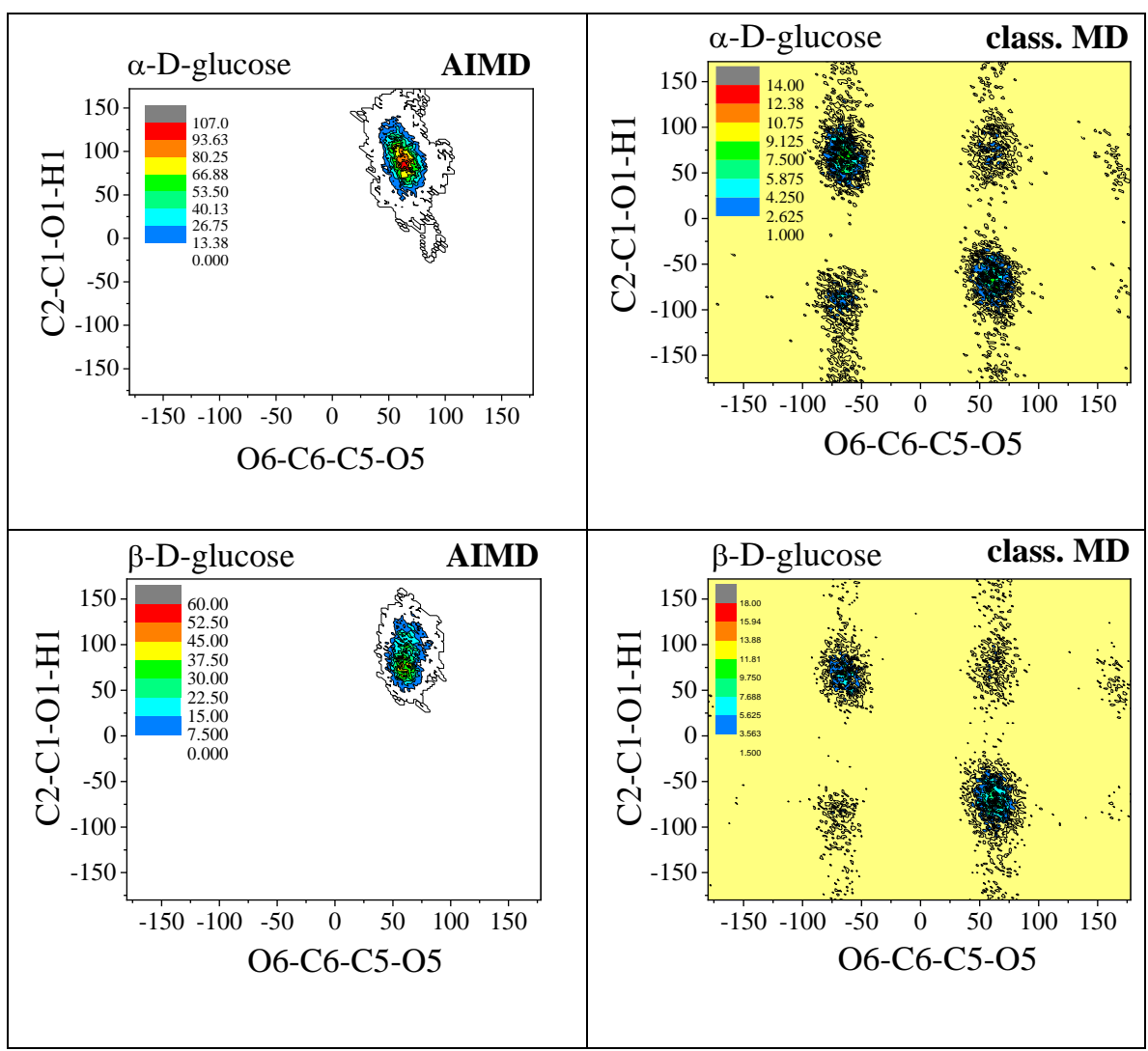

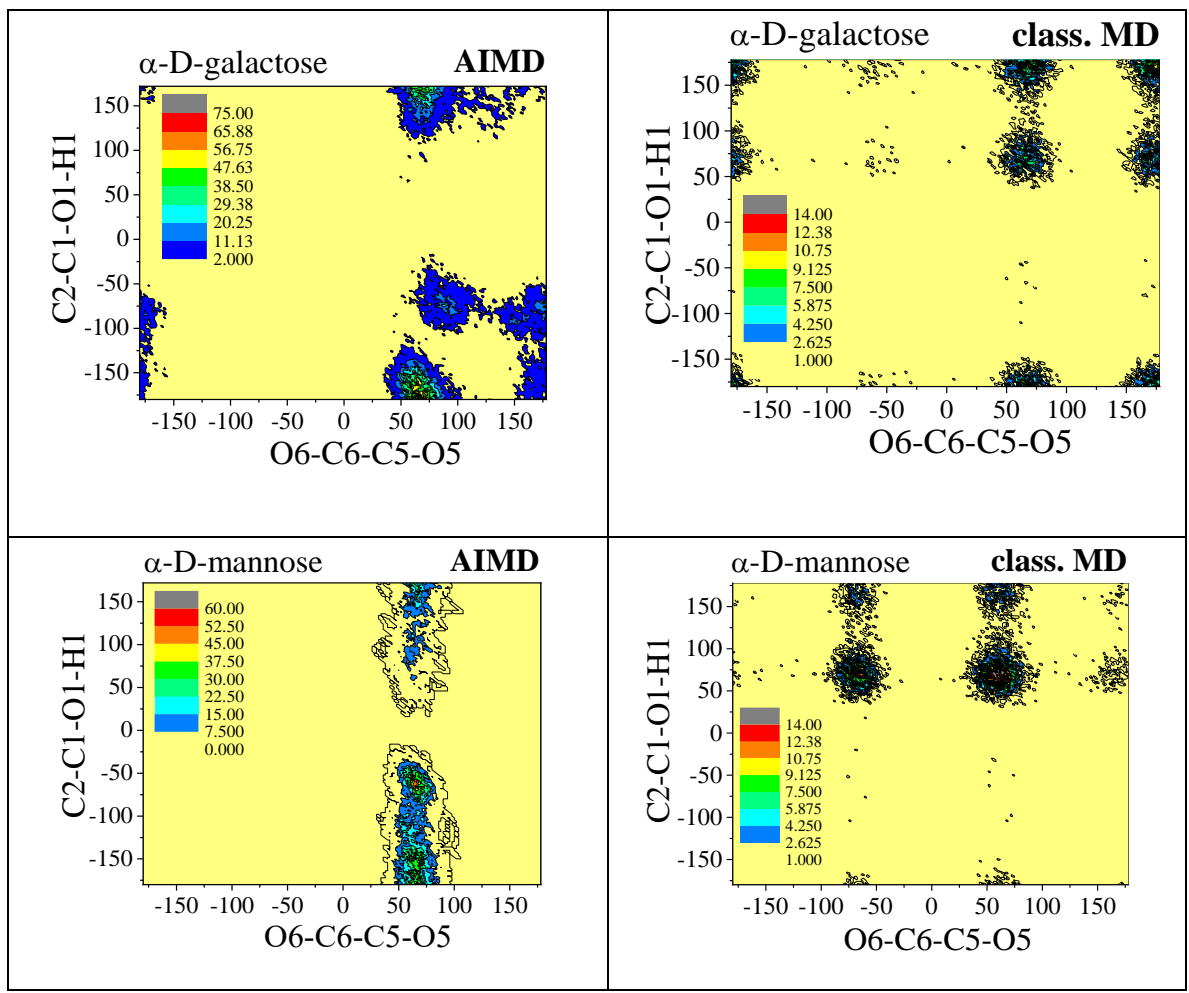

**Figure S2.** Two-dimensional contour plots of C2-C1-O1-H1 and O6-C5-C5-O5 dihedral angles.

## 4. Partial radial distribution functions

Figure S3 shows the partial radial distribution functions between the sugar and water hydroxyl groups, namely $H_i$-$O_{water}$ (i=1,2,3,4,5) and $O_i$-$H_{water}$ (i=1,2,3,4,5,6) for β-D-glucose.

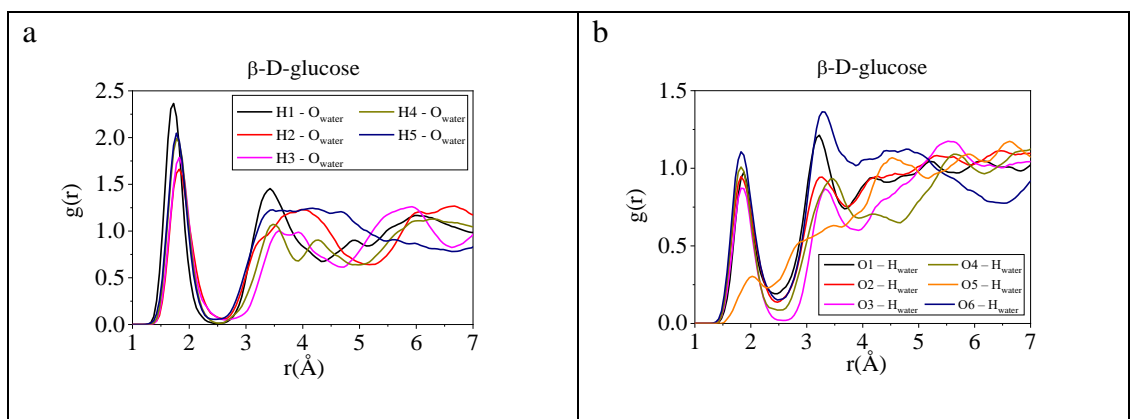

**Figure S3.** (a) The $H_i$-$O_{water}$ (i=1,2,3,4,5) partial radial distribution functions for β-D-glucose. (b) The $O_i$-$H_{water}$ (i=1,2,3,4,5,6) partial radial distribution functions for β-D-glucose.

The $C_i$-$O_{water}$ (i=1, 2, 3, 4, 5, 6) radial distributions are presented in Figure S4 for β-D-glucose.

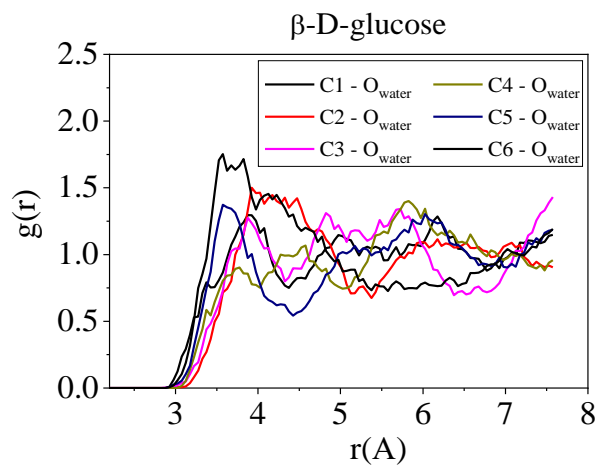

**Figure S4**. $C_i$-$O_{water}$ partial radial distribution functions.

## 5. Cooperative behavior of water molecules around the sugar molecule

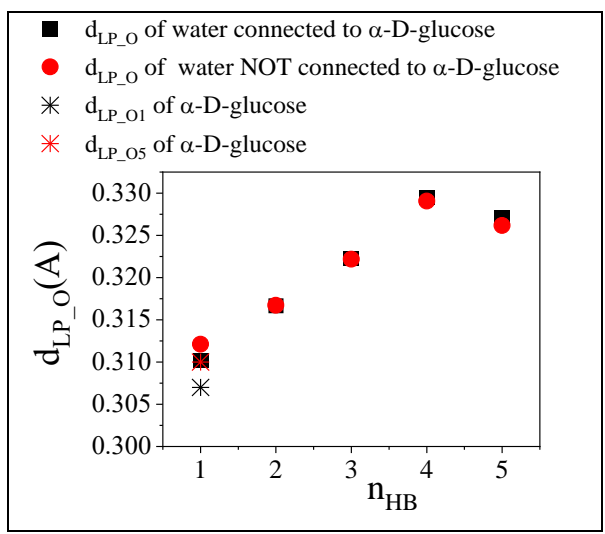

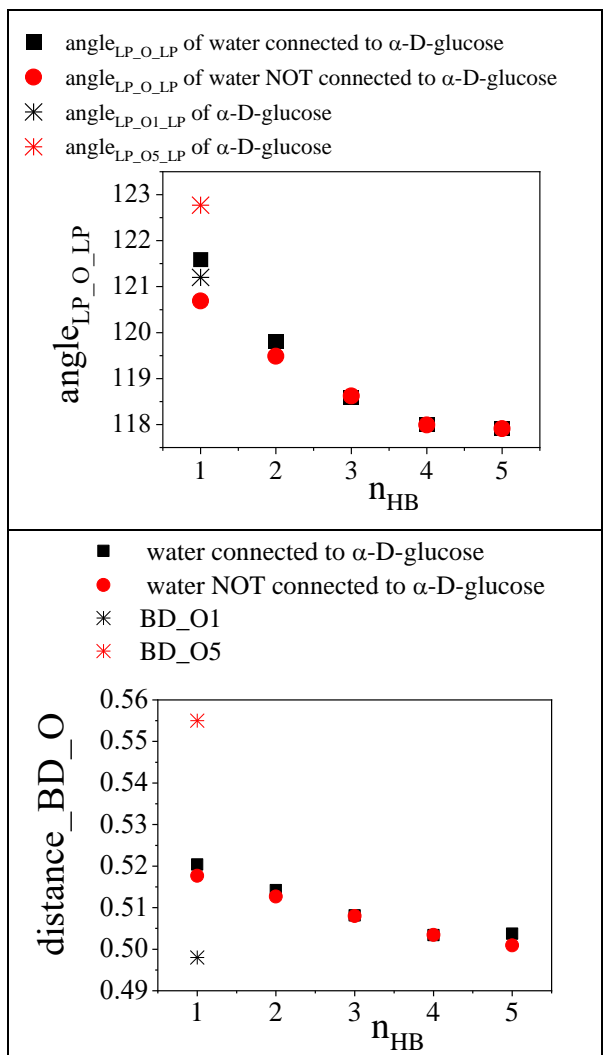

**Figure S5.** Characteristic distances and angles related to the centers of localized orbital of O atoms (of water, monosaccharide).

In water molecules, the four localized (hydrogen-bonding) orbitals lie in the directions of the two lone pairs of O atoms (LP_O), and of the two O-H vectors (BD_O). For monosaccharides, these localized orbitals can be found in the directions of the lone pairs of $O_{sugar}$. $O_{sugar}$-H and $O_{sugar}$-$C_{sugar}$ directions. We calculated the average distance from the oxygen atoms to the centers of the lone pairs of O-atoms of water molecules ($d_{LP\_O}$) that are, and also, those that are *not*, connected to monosaccharides. Fig. S5 shows the average distance from the O1 ($d_{LP\_O1}$) and O5 ($d_{LP\_O5}$) sites to the centres of the lone pairs for α-D-glucose. $d_{LP\_O}$ increases almost linearly with the number of hydrogen-bonded neighbours, which suggests an increasing interaction strength (since the lone pair gets a little closer to the neighboring atom). Similar behavior was detected in pure liquid water. [3] $d_{LP\_O1}$ and $d_{LP\_O5}$ (referring to O atoms in saccharide molecules) are shorter than the shortest $d_{LP\_O}$ d (O atoms in water molecules) distance. 'distance_BD_O' (in the bottom part of Fig. S5) refers to the distance between an O atom and the center of the (localized) electronic density in the direction of the covalently bonded neighbour (H in water, and H or C in the saccharide). Note that 'distance_BD_O' also varies systematically with the number of H-bonded neighbours.

**References:**

[1] J. N. Chythra, S. S. Mallajosyula, Impact of Polarization on the Ring Puckering Dynamics of Hexose Monosaccharides, J. Chem. Inf. Model. 63 (2023) 208-223, https://doi.org/10.1021/acs.jcim.2c01286.

[2] H. S. Hansen, P. H. Hünenberger, A reoptimized GROMOS force field for hexopyranose-based carbohydrates accounting for the relative free energies of ring conformers, anomers, epimers, hydroxymethyl rotamers, and glycosidic linkage conformers, J Comput. Chem. 32 (2011) 998–1032, https://doi.org/10.1002/jcc.21675.

[3] I. Tvaroška, F. R. Taravel, J. P. Utille, J. P. Carver, Quantum Mechanical and NMR Spectroscopy Studies on the Conformations of the Hydroxymethyl and Methoxymethyl Groups in Aldohexosides, Carbohydr. Res. 337 (2002) 353−367, https://doi.org/10.1016/S0008-6215(01)00315-9.